\documentclass[aps,prl,twocolumn,superscriptaddress,showpacs,preprintnumbers,amsmath,amssymb]{revtex4}

\usepackage{graphicx,color} % Include figure files
\usepackage{dcolumn}  % Align table columns on decimal point

\graphicspath{{ps}}

\begin{document}

%\vspace*{-3\baselineskip}
%\resizebox{!}{3cm}{\includegraphics{belle.eps}}

\preprint{\vbox{ \hbox{   }
}}

\title{ \quad\\[0.5cm]  Search for lepton-flavor-violating 
$\tau \to \ell V^0$ decays at Belle}
%%%% >>>>> insert the authorlist here. BEFORE the abstract !!!!! <<<<<

\affiliation{Budker Institute of Nuclear Physics, Novosibirsk, Russia}
\affiliation{Chiba University, Chiba, Japan}
\affiliation{University of Cincinnati, Cincinnati, OH, USA}
\affiliation{The Graduate University for Advanced Studies, Hayama, Japan}
\affiliation{Hanyang University, Seoul, South Korea}
\affiliation{University of Hawaii, Honolulu, HI, USA}
\affiliation{High Energy Accelerator Research Organization (KEK), Tsukuba, Japan}
\affiliation{Institute of High Energy Physics, Chinese Academy of Sciences, Beijing, PR China}
\affiliation{Institute for High Energy Physics, Protvino, Russia}
\affiliation{Institute of High Energy Physics, Vienna, Austria}
\affiliation{Institute for Theoretical and Experimental Physics, Moscow, Russia}
\affiliation{J. Stefan Institute, Ljubljana, Slovenia}
\affiliation{Kanagawa University, Yokohama, Japan}
\affiliation{Korea University, Seoul, South Korea}
\affiliation{Kyungpook National University, Taegu, South Korea}
\affiliation{\'Ecole Polytechnique F\'ed\'erale de Lausanne, EPFL, Lausanne, Switzerland}
\affiliation{Faculty of Mathematics and Physics, University of Ljubljana, Ljubljana, Slovenia}
\affiliation{University of Maribor, Maribor, Slovenia}
\affiliation{University of Melbourne, Victoria, Australia}
\affiliation{Nagoya University, Nagoya, Japan}
\affiliation{Nara Women's University, Nara, Japan}
\affiliation{National Central University, Chung-li, Taiwan}
\affiliation{National United University, Miao Li, Taiwan}
\affiliation{Department of Physics, National Taiwan University, Taipei, Taiwan}
\affiliation{H. Niewodniczanski Institute of Nuclear Physics, Krakow, Poland}
\affiliation{Nippon Dental University, Niigata, Japan}
\affiliation{Niigata University, Niigata, Japan}
\affiliation{University of Nova Gorica, Nova Gorica, Slovenia}
\affiliation{Osaka City University, Osaka, Japan}
\affiliation{Osaka University, Osaka, Japan}
\affiliation{Panjab University, Chandigarh, India}
\affiliation{Saga University, Saga, Japan}
\affiliation{University of Science and Technology of China, Hefei, PR China}
\affiliation{Seoul National University, Seoul, South Korea}
\affiliation{Sungkyunkwan University, Suwon, South Korea}
\affiliation{University of Sydney, Sydney, NSW, Australia}
\affiliation{Tata Institute of Fundamental Research, Mumbai, India}
\affiliation{Toho University, Funabashi, Japan}
\affiliation{Tohoku Gakuin University, Tagajo, Japan}
\affiliation{Tohoku University, Sendai, Japan}
\affiliation{Department of Physics, University of Tokyo, Tokyo, Japan}
\affiliation{Tokyo Metropolitan University, Tokyo, Japan}
\affiliation{Tokyo University of Agriculture and Technology, Tokyo, Japan}
\affiliation{Virginia Polytechnic Institute and State University, Blacksburg, VA, USA}
\affiliation{Yonsei University, Seoul, South Korea}
\author{Y.~Nishio} % Nagoya 
\affiliation{Nagoya University, Nagoya, Japan}
\author{K.~Inami} % Nagoya 
\affiliation{Nagoya University, Nagoya, Japan}
\author{T.~Ohshima} % Nagoya 
\affiliation{Nagoya University, Nagoya, Japan}
\author{I.~Adachi} % KEK 
\affiliation{High Energy Accelerator Research Organization (KEK), Tsukuba, Japan}
\author{H.~Aihara} % Tokyo 
\affiliation{Department of Physics, University of Tokyo, Tokyo, Japan}
\author{K.~Arinstein} % BINP 
\affiliation{Budker Institute of Nuclear Physics, Novosibirsk, Russia}
\author{V.~Aulchenko} % BINP 
\affiliation{Budker Institute of Nuclear Physics, Novosibirsk, Russia}
\author{T.~Aushev} % Lausanne  ITEP 
\affiliation{\'Ecole Polytechnique F\'ed\'erale de Lausanne, EPFL, Lausanne, Switzerland}
\affiliation{Institute for Theoretical and Experimental Physics, Moscow, Russia}
\author{T.~Aziz} % Tata 
\affiliation{Tata Institute of Fundamental Research, Mumbai, India}
\author{A.~M.~Bakich} % Sydney 
\affiliation{University of Sydney, Sydney, NSW, Australia}
% ...
\collaboration{The Belle Collaboration}
\noaffiliation

%\collaboration{Belle Collaboration}
%\noaffiliation

\begin{abstract}
We have searched for neutrinoless $\tau$ lepton decays into $\ell$ and
$V^0$, where $\ell$ stands for an electron or muon, and 
$V^0$ for a vector meson ($\phi$, $\omega$, $K^{*0}$, $\bar{K}^{*0}$ 
or $\rho^0$), using 543~fb$^{-1}$ of data collected
with the Belle detector at the KEKB asymmetric-energy $e^+e^-$ collider.
No excess of signal events over the expected
background has been observed, and we set upper limits on the 
branching fractions in the range $(5.9-18) \times 10^{-8}$ at the 
90\% confidence level. These upper limits 
include the first results for the $\ell \omega$ mode
as well as new limits that are
significantly more restrictive than our previous results for the 
$\ell \phi$, $\ell K^{*0}$, $\ell \bar{K}^{*0}$ and $\ell \rho^0$ modes.
\end{abstract}


\maketitle

%%%% >>>> keep the final version single-spaced
\tighten

\section{Introduction}

In the Standard Model (SM), lepton-flavor-violating (LFV) decays of charged
leptons are forbidden;
even if neutrino mixing is taken into account, they are highly suppressed.
However, LFV is expected to appear in many extensions of the SM.
Some such models predict branching fractions
for $\tau$ LFV decays in the range $10^{-8}-10^{-7}$~\cite{Amon,BR,CG},
which can be reached at the present B-factories.
An observation of LFV would provide unambiguous evidence for new physics 
beyond the SM.

A search for LFV $\tau^-$ decays into neutrinoless 
final states with one 
charged lepton $\ell^-$($e^-$ or $\mu^-$) and a vector meson
was first performed by the CLEO collaboration in the 
$\ell^- \phi$, $\ell^- K^{*0}$, $\ell^- \bar{K}^{*0}$ and 
$\ell^- \rho^0$ final states in 1998; this search set upper
limits in the range $(2.0 - 7.5) \times 10^{-6}$ ~\cite{CLEO}.
Later, Belle obtained upper limits in the range 
$(2.0 - 7.7) \times 10^{-7}$ using 158 fb$^{-1}$
of data.
In this paper, we report an improved search for 
LFV $\tau^-$ decays~\cite{CC}
into a charged lepton and a neutral vector meson ($V^0$), where $V^0$ includes
$\omega$ in addition to the $\phi$, $K^{*0}$, $\bar{K}^{*0}$ and $\rho^0$
\footnotemark[2]
\footnotetext[2]{While preparing this paper, we became aware that
the BaBar group had also reported 
on a search for $\tau^- \to \ell^- \omega$ 
in a preprint~\cite{BaBar}.}.
The analysis is based on a data sample of 543 fb$^{-1}$,
corresponding to $4.99\times10^8$ $\tau$-pairs collected with the Belle 
detector~\cite{Belle} at the KEKB asymmetric-energy $e^+e^-$
collider~\cite{KEKB} taken at the $\Upsilon(4S)$ resonance and 60 MeV below it.
These results supersede our previous published results~\cite{previous_Belle}.

The Belle detector is a large-solid-angle magnetic spectrometer that
consists of a silicon vertex detector,
a 50-layer central drift chamber, an array of
aerogel threshold Cherenkov counters, 
a barrel-like arrangement of time-of-flight
scintillation counters, and an electromagnetic calorimeter
comprised of CsI(Tl) crystals located inside 
a superconducting solenoid coil that provides a 1.5~T
magnetic field.  An iron flux-return located outside 
the coil is instrumented to detect $K_L^0$ mesons and to identify
muons.  The detector is described in detail elsewhere~\cite{Belle}.
Two inner detector configurations were used. A 2.0 cm radius beam-pipe
and a 3-layer silicon vertex detector were used for the first sample
of 158~fb$^{-1}$, while a 1.5 cm radius beam-pipe, a 4-layer
silicon detector and a small-cell inner drift chamber were used to record  
the remaining 385~fb$^{-1}$~\cite{svd2}.  

\section{Event Selection}

We search for 
%$\tau \to \ell \phi$, $\ell \omega$, $\ell K^{*0}$,
%$\ell \bar{K}^{*0}$ and $\ell \rho^0$ 
events, in 
which one $\tau$ decays to a charged lepton and
two charged hadrons (3-prong decay) while the other $\tau$ decays 
into one charged particle (1-prong decay) and missing 
neutral particle(s).
%allowed within SM.
We reconstruct $\phi$ candidates from $K^+ K^-$,
$\omega$ from $\pi^+\pi^-\pi^0$,
$K^{*0}$ from $K^+ \pi^-$, $\bar{K}^{*0}$ from $K^- \pi^+$
and $\rho^0$ from $\pi^+\pi^-$.

The selection criteria described below are optimized 
from studies of Monte Carlo (MC) simulated events
and the experimental data distributions. The background
estimation is based on MC simulations
of the reaction $e^+e^- \to \tau^+\tau^-$ as well as
$q\bar{q}$ continuum and two-photon processes. The $\tau^+\tau^-$ sample
corresponding to 1524 fb$^{-1}$ is generated using the KKMC code~\cite{KKMC}.
The MC samples of $q \bar{q}$ and two-photon processes are produced using  
EvtGen~\cite{EvtGen} and AAFH~\cite{AAFH}, respectively,
in amounts corresponding to the luminosity of the experiment.
The signal MC events are generated by KKMC assuming a phase-space distribution
for $\tau$ decay.
The detector response is simulated by a GEANT3~\cite{geant} based program.

The transverse momentum for each charged track is required to be larger than 
0.06 GeV/$c$ in the barrel region ($-0.6235<\cos \theta<0.8332$, where 
$\theta$ is the polar angle relative to the direction opposite to that of
the incident $e^+$ beam in the laboratory frame)
and 0.1 GeV/$c$ in the endcap region ($-0.8660<\cos \theta<-0.6235$ and 
$0.8332<\cos \theta<0.9563$).
The energies of photon candidates are required to be 
larger than 0.1 GeV in both regions. 

To select the signal topology,
we require four charged tracks in an event with zero net charge,
and the total energy of charged tracks and photons
in the center-of-mass (CM) frame to be less than 11 GeV.
We also require
the missing momentum in the laboratory frame to exceed
0.6 GeV/$c$, and to point into the
detector acceptance ($-0.8660 < \cos \theta < 0.9563$).
Here the missing momentum is defined as the 
difference between the momentum of the initial $e^+e^-$
system, and the sum of the observed momentum vectors.
An event is subdivided into 3-prong and 1-prong
hemispheres with respect to the thrust axis calculated
from the momenta of all charged tracks and photons 
in the CM frame.
These hemispheres are referred to as the signal and tag sides, respectively.
We allow at most two photons on the tag side
to take into account initial state radiation.
To reduce the $q\bar{q}$ background, not more than one photon 
on the signal side
is allowed for the $\ell^- \phi$, $\ell^- K^{*0}$, $\ell^- \bar{K}^{*0}$,
$\ell^- \rho^0$ modes while not more than two photons
in addition to $\pi^0$ daughters are permitted for the $\ell^- \omega$ 
modes.
A charged particle of the type $x$ ($x = \mu$, $e$, $K$ or $\pi$)
is identified using the likelihood ratio parameter, $P_{x}$. 
This is defined as $P_x = L_x/(\sum_x L_x)$, 
where $L_x$ is the likelihood for particle type $x$,
determined from the responses of the relevant detectors~\cite{PID}.
For muon candidates on the signal side we require 
$P_{\mu} >$ 0.95
while their momentum should be greater than 1.0 GeV/$c$. 
The efficiency for muon identification is 92\% 
with a 1.2\% probability to misidentify a pion as a muon.
Electrons on the signal side are required to have $P_e>0.9$ 
and momenta greater than 0.5 GeV/$c$.
The efficiency for the electron identification is
94\% while the probability to misidentify a pion as an electron
is 0.1\%.

Candidate $\phi$ mesons are selected from  $K^+ K^-$ pairs
with invariant mass in the range
$1.01~{\rm GeV}/c^2<M_{K^+K^-}<1.03~{\rm GeV}/c^2$~($\pm 4\sigma$).
For both kaon daughters we require
$P_K > 0.8$. 
To reduce the background from the $\gamma \to e^+ e^-$ 
conversions the cut $P_e < 0.1$ is applied.

Candidate $\omega$ mesons are reconstructed from $\pi^+\pi^-\pi^0$
with the invariant mass requirement 
$0.757~{\rm GeV}/c^2<M_{\pi^+\pi^-\pi^0}<0.808~{\rm GeV}/c^2$~($\pm 3\sigma$). 
A $\pi^0$ candidate is selected from $\gamma$ pairs with invariant mass 
in the range $0.11~{\rm GeV}/c^2<M_{\gamma \gamma}<0.15~{\rm GeV}/c^2$.
In order to improve the $\omega$ mass resolution,
the $\pi^0$ mass is constrained to its world average value of
 $134.9766~{\rm GeV}/c^2$
for the $\omega$ mass reconstruction.

Candidate $K^{*0}$ and $\bar{K}^{*0}$ mesons are selected from
$K^{\pm}\pi^{\mp}$ pairs with invariant mass in the range
$0.827~{\rm GeV}/c^2<M_{K\pi}<0.968~{\rm GeV}/c^2$~($\pm 3\sigma$),
which satisfy the condition $P_K>0.8$ for the kaon candidate
and $P_e <0.1$ for both daughters.

Candidate $\rho^0$ mesons are selected from $\pi^{+}\pi^{-}$ pairs with
invariant mass in the range
$0.478~{\rm GeV}/c^2<M_{\pi^+\pi^-}<1.074~{\rm GeV}/c^2$ ~($\pm 4\sigma$),
requiring that the daughter pions have $P_K < 0.1$, $P_e < 0.1$ 
and momenta greater than 0.5 GeV/$c$.  
In addition, for the $\tau^-\to e^-\rho^0$ mode, we
require $P_\mu < 0.5$ for daughter pions 
in order to reduce the two-photon background from $ee \to ee\mu\mu$.

\begin{figure}[htb]
\includegraphics[height=0.23\textwidth]{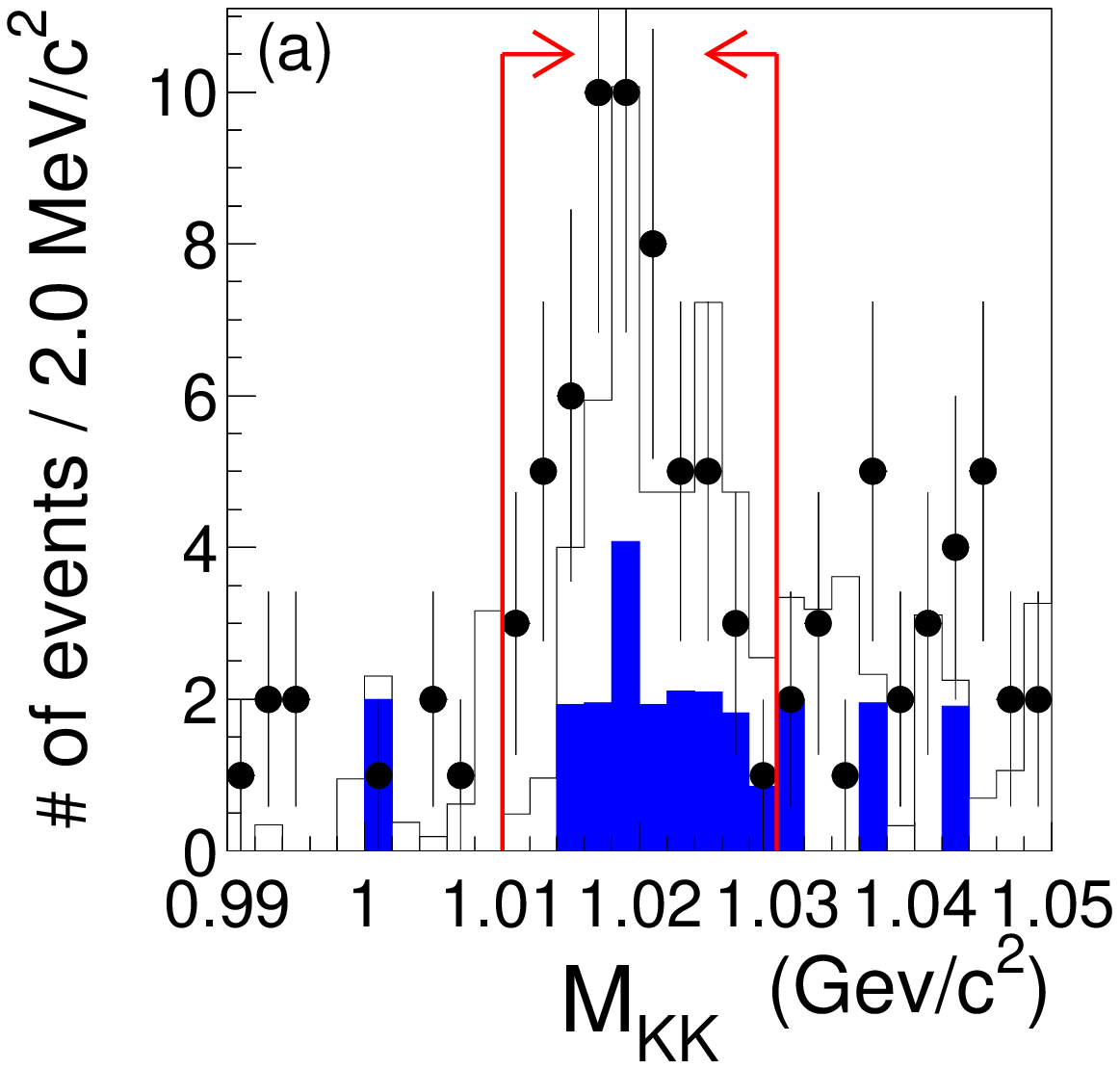}
\includegraphics[height=0.23\textwidth]{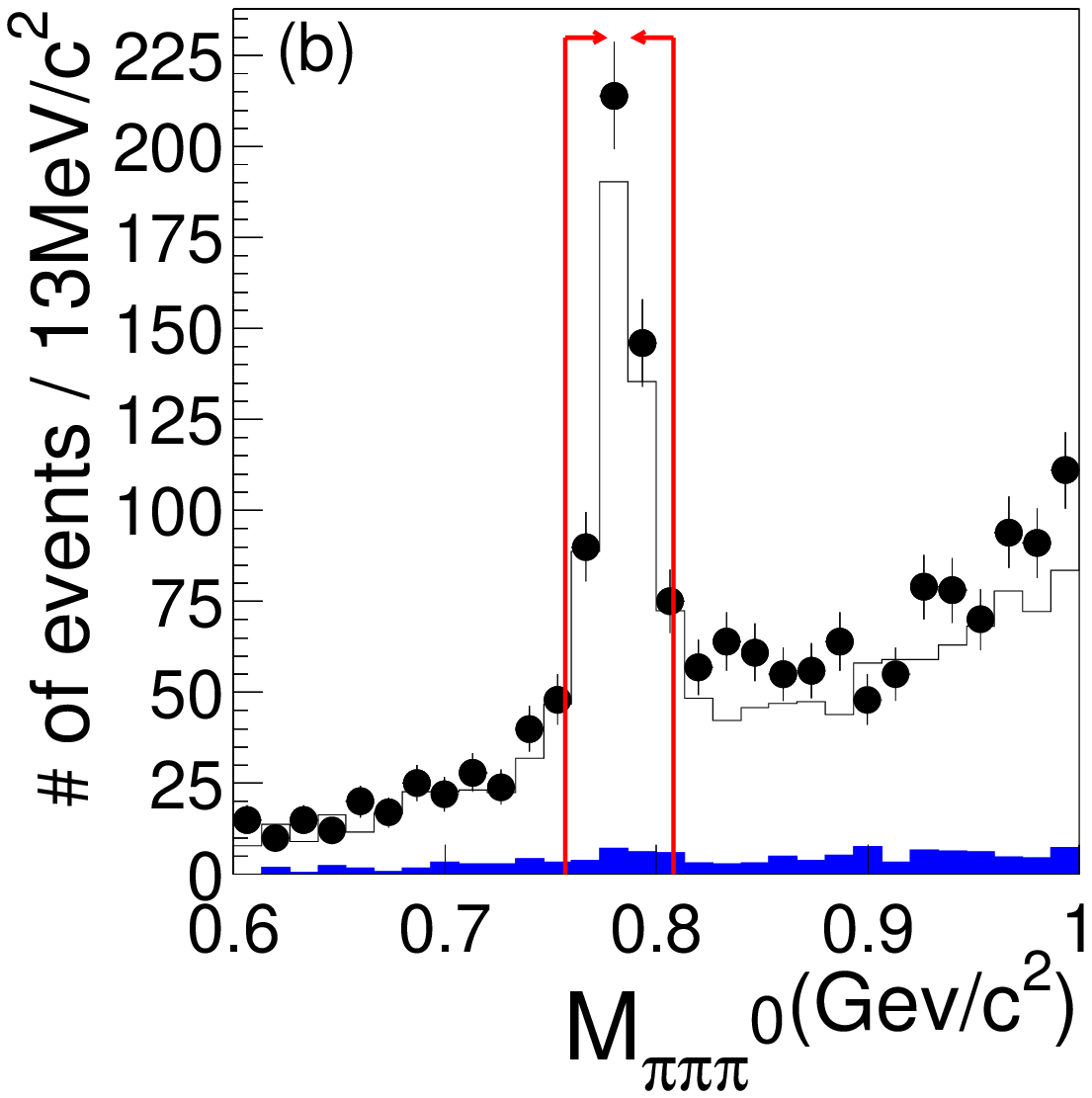}
\includegraphics[height=0.23\textwidth]{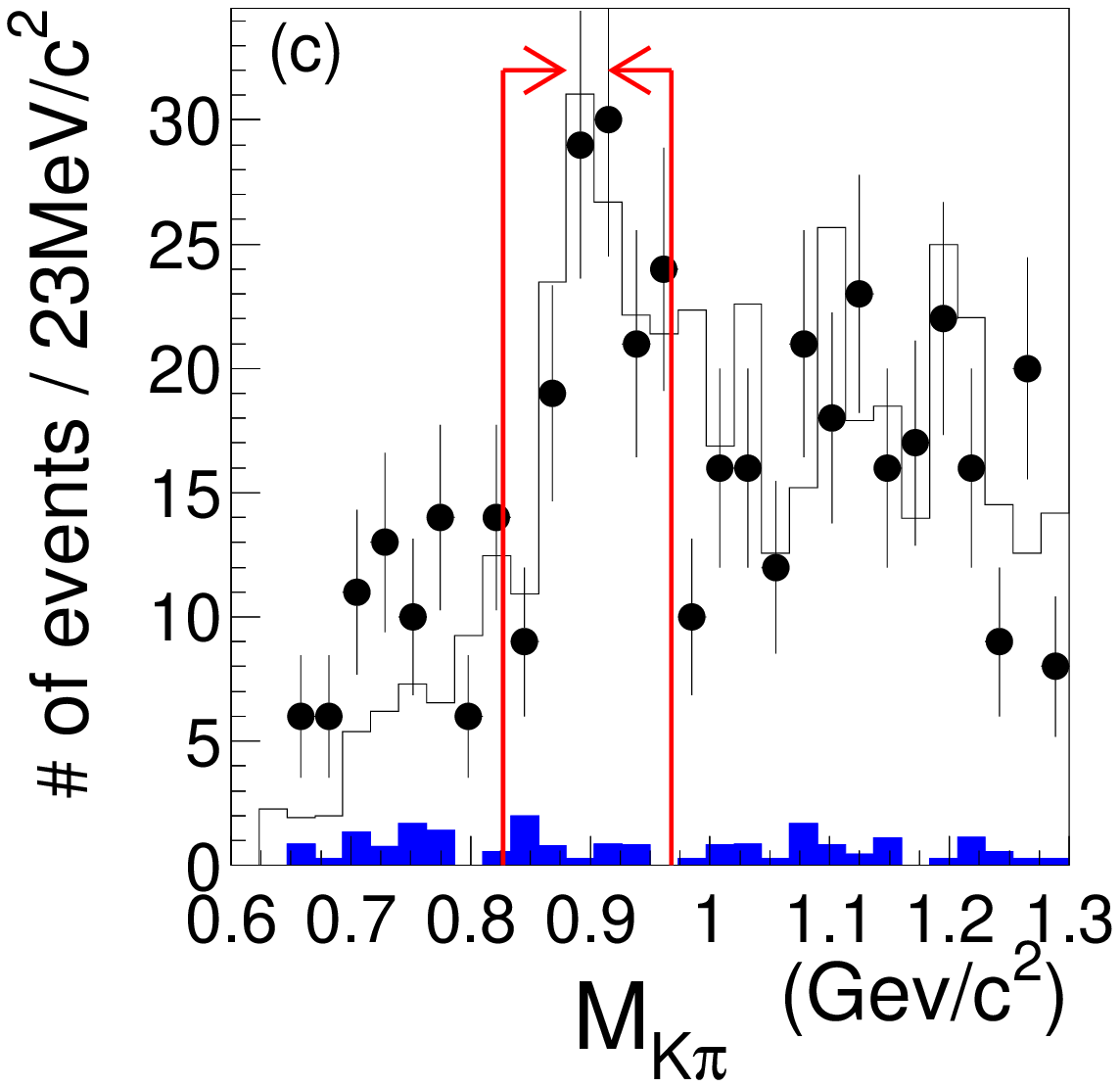}
\includegraphics[height=0.23\textwidth]{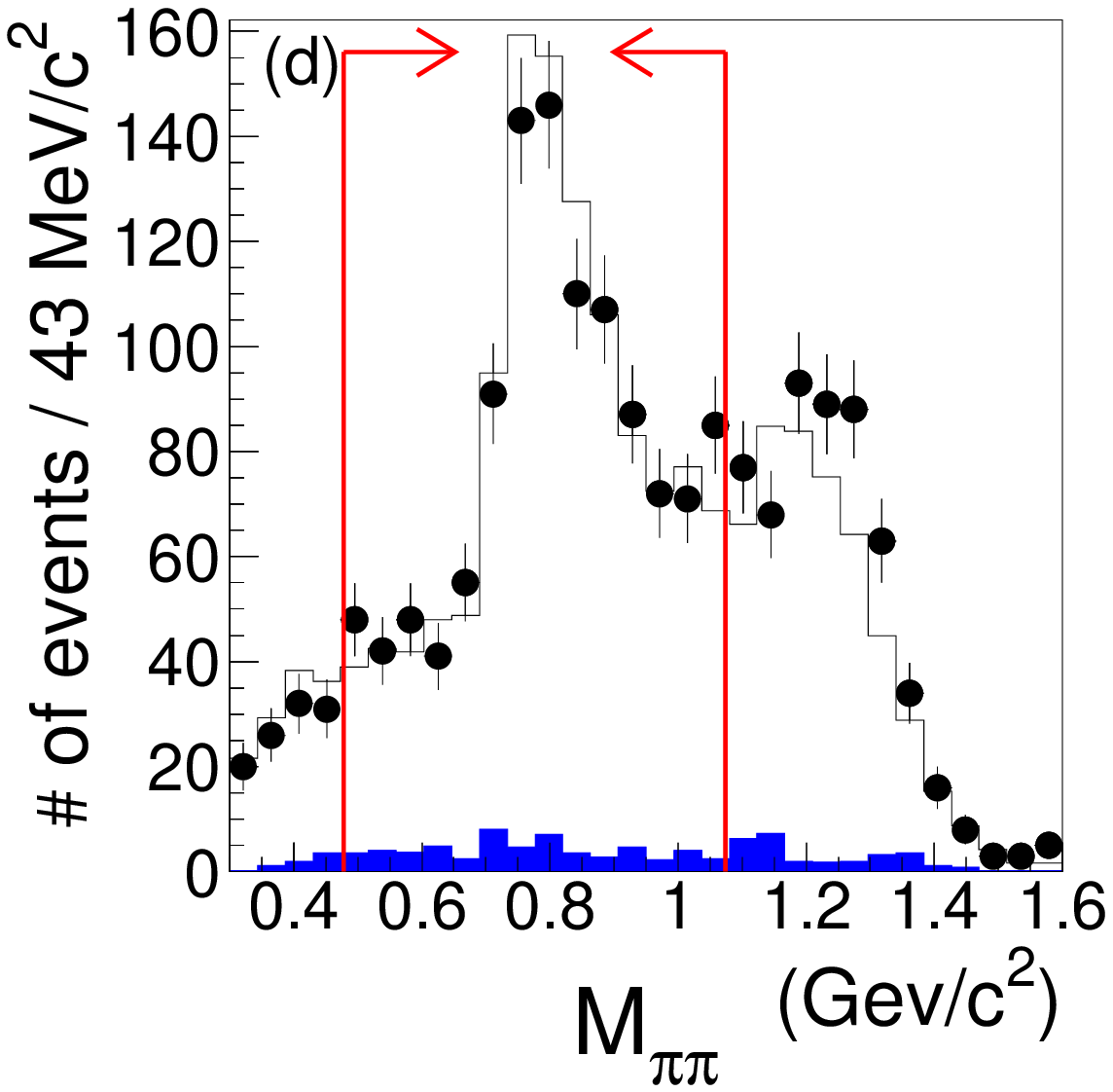}
\caption{Invariant mass distributions of
(a) $\phi \to K^+ K^-$ for $\tau^- \to \mu^- \phi$,
(b) $\omega \to \pi^+ \pi^- \pi^0$ for $\tau^- \to \mu^- \omega$,
(c) $K^{*0} \to K^+ \pi^-$ for $\tau^- \to \mu^- K^{*0}$ and
(d) $\rho^0 \to \pi^+ \pi^-$ for $\tau^- \to \mu^- \rho^0$
in the region $1.5$ GeV/$c^2<M_{\ell V^0}<1.95$ GeV/$c^2$ and $-0.5$ GeV$
< \Delta E<0.5$ GeV.
The points with error bars are data.
The open histogram shows the expected $\tau^+\tau^-$ background MC
while the filled histogram is the sum of $q\bar{q}$ and two-photon
MCs. The regions between the vertical lines are selected.}
\label{fig:Vmass}
\end{figure}

Figures~\ref{fig:Vmass}~(a)-(d) show the invariant mass distributions of
the $\phi$, $\omega$, $K^{*0}$ and $\rho^0$ candidates for the 
$\tau^- \to \mu^- \phi$,
$\tau^- \to \mu^- \omega$, $\tau^- \to \mu^- K^{*0}$ and 
$\tau^- \to \mu^- \rho^0$ modes, respectively.
The estimated background distributions agree with the data.
The main background contribution for the $\tau^- \to \ell^- \phi$ mode 
is due to $q\bar{q}$ events involving $\phi$ mesons.
For the $\tau^- \to \ell^- \omega$ mode 
the dominant background comes from
$\tau^- \to \pi^- \omega \nu_{\tau}$ decay with 
the pion misidentified as a lepton.
The $\tau^- \to \pi^- \pi^+ \pi^- \nu_{\tau}$ decay is one of
the main background sources for the 
$\tau^- \to \ell^- K^{*0}$, $\ell^- \bar{K}^{*0}$ and 
$\tau^- \to \ell^- \rho^0$ modes.
In this background source one pion is misidentified as a lepton
for all modes while for the $\tau^- \to \ell^- K^{*0}$ and 
$\ell^- \bar{K}^{*0}$ modes one additional pion is
misidentified as a kaon.
          
\begin{figure}
\begin{center}
\includegraphics[height=0.23\textwidth]{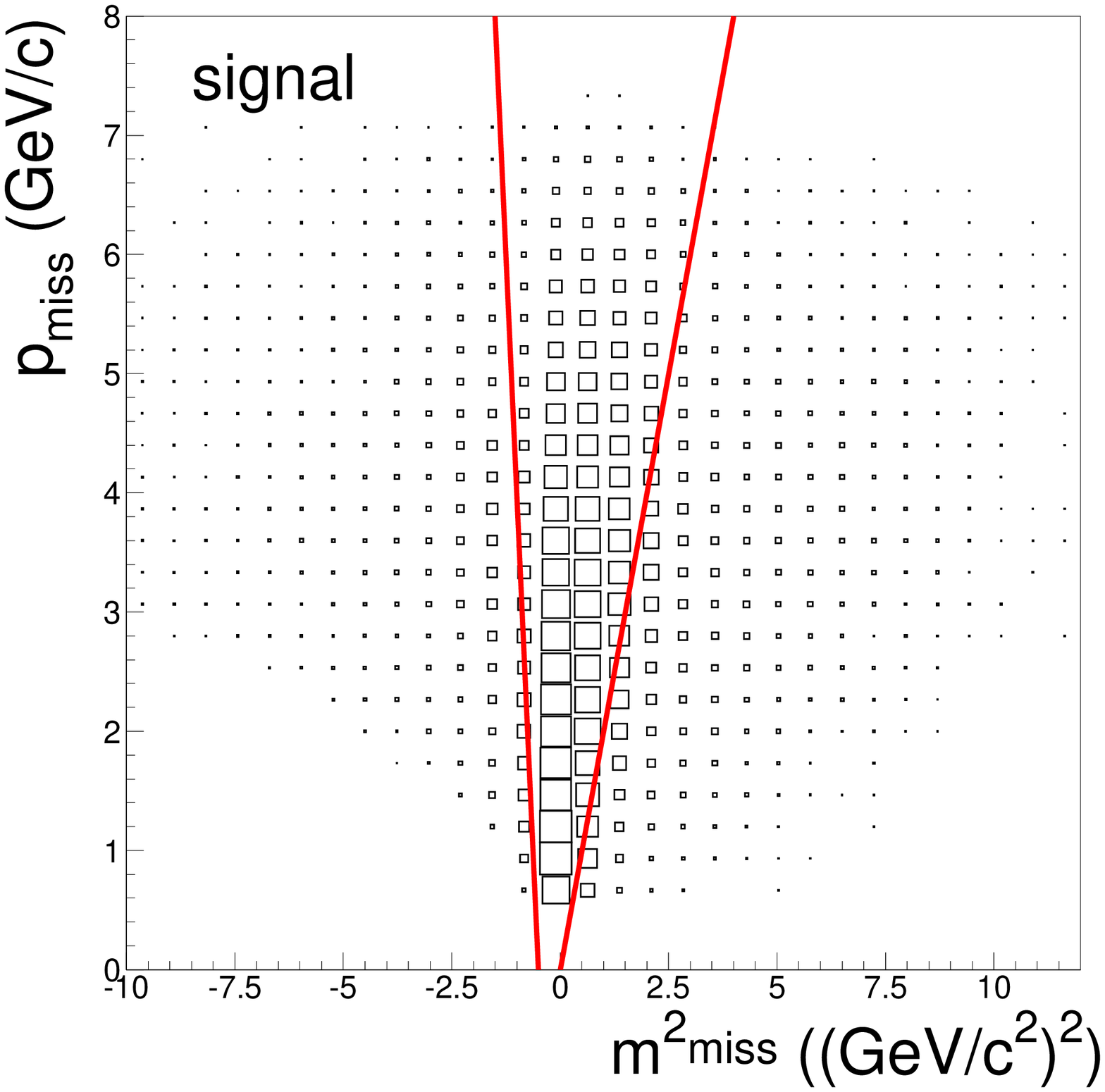}
\includegraphics[height=0.23\textwidth]{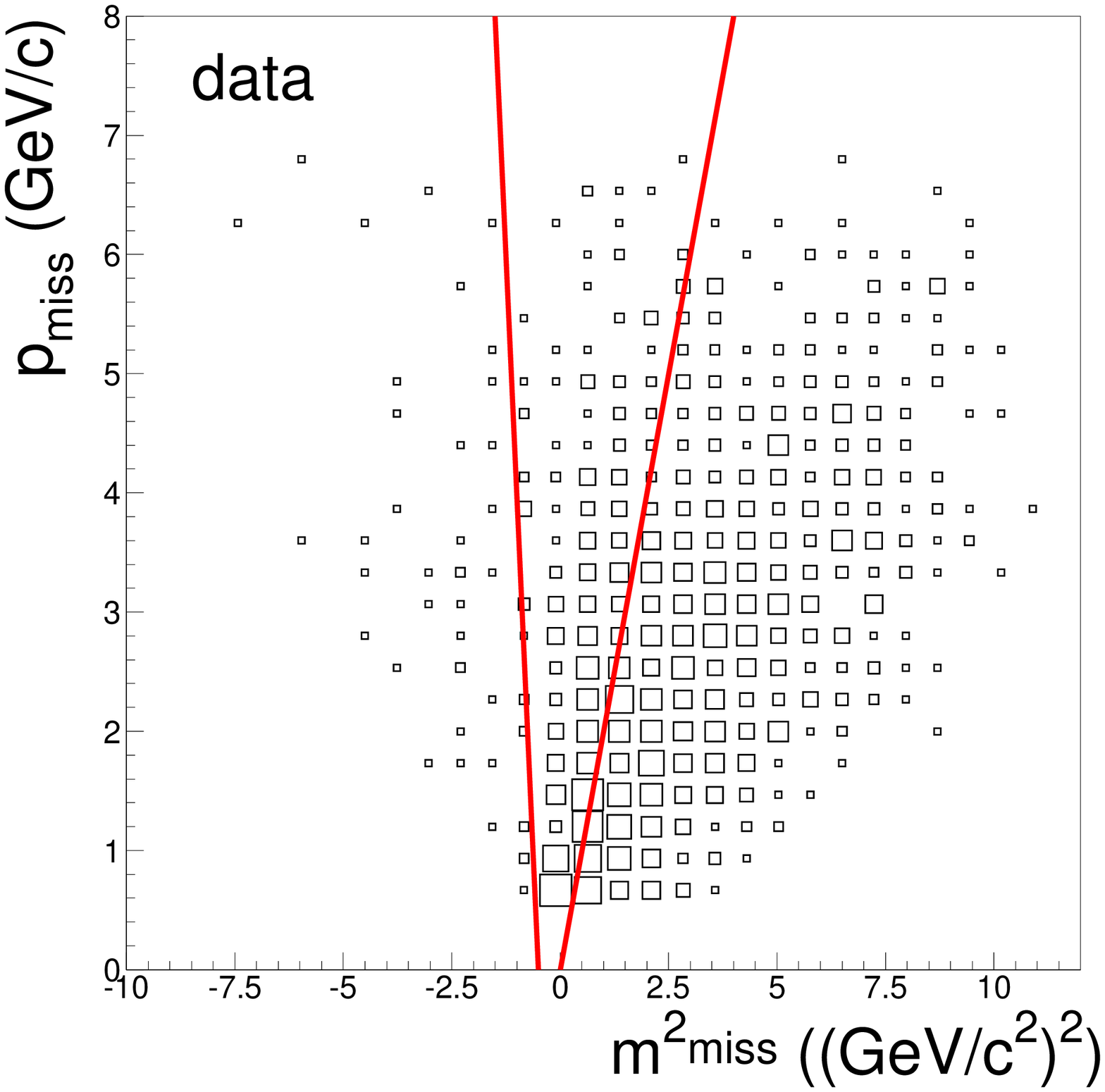}
\end{center}
\caption{$p_{\rm miss}$ vs. $m^2_{\rm miss}$ plots 
for signal MC and data for the $\tau \to \mu \rho^0$ mode. 
The regions between the vertical lines are selected.}
\label{fig:vcut}
\end{figure}

To reduce the remaining background from $\tau^+\tau^-$ and $q\bar{q}$,
the events from the triangular area defined by
the missing momentum, $p_{\rm miss}$ (GeV/$c$),
and missing mass squared, $m^2_{\rm miss}$ ((GeV/$c^2$)$^2$)
are selected for further consideration.
These requirements are summarized 
in Table~\ref{tbl:vcut} and illustrated in 
Fig.~\ref{fig:vcut} by the two-dimensional plots of $p_{\rm miss}$ (GeV/$c$)  
versus $m^2_{\rm miss}$ ((GeV/$c^2$)$^2$) for the $\tau^-\to\mu^- \rho^0$ mode.
\begin{table}
\caption{Selection criteria using $p_{\rm miss}$ (GeV/$c$) 
and $m^2_{\rm miss}$ ((GeV/$c^2$)$^2$) where $p_{\rm miss}$ is 
 missing momentum and $m^2_{\rm miss}$ is missing mass squared.}
\label{tbl:vcut}
\begin{center}
\begin{tabular}{lll} \hline
Mode &  ~~Selection criteria  \\ 
$\tau^-\to$    & ~~                       \\ \hline
 $\ell^- \phi$ & ~~$p_{\rm miss} > \frac{8}{9} m^2_{\rm miss}$ & and $m^2_{\rm miss}> -0.5$ \\ 
 $\ell^- \omega$ & ~~$p_{\rm miss} > \frac{8}{3} m^2_{\rm miss} -\frac{8}{3} $ & and $m^2_{\rm miss}> -0.5$  \\ 
 $\mu^- K^{*0}$  & ~~$p_{\rm miss} > \frac{8}{4.5} m^2_{\rm miss}-\frac{8}{9}$ & and $p_{\rm miss} > 8 m^2_{\rm miss}$ \\
 $e^- K^{*0}$  & ~~$p_{\rm miss} > \frac{8}{5.5} m^2_{\rm miss}-\frac{8}{11}$ & and $m^2_{\rm miss}>0$ \\
 $\mu^- \bar{K}^{*0}$  & ~~$p_{\rm miss} > \frac{8}{6.5} m^2_{\rm miss}$ & and $m^2_{\rm miss}>-0.5$ \\
 $e^- \bar{K}^{*0}$  & ~~$p_{\rm miss} > \frac{6}{5} m^2_{\rm miss}$ & and $p_{\rm miss} > -\frac{8}{1.4} m^2_{\rm miss}$ \\ 
 $\mu^- \rho^0$  & ~~$p_{\rm miss} > -8 m^2_{\rm miss}-4$ & and $p_{\rm miss}>2m^2_{\rm miss}$ \\
 $e^- \rho^0$  & ~~$p_{\rm miss} > -8 m^2_{\rm miss}-4$ & and $p_{\rm miss} > 1.6 m^2_{\rm miss}$ \\ 
\hline
\end{tabular}
\end{center}
\end{table}

For the $\ell^- \omega$ ($\ell^- K^{*0}$ and $\ell^- \bar{K}^{*0}$) mode, 
we require that the opening angle between 
the lepton and $\omega$ ($K^{*0}$) on the signal side in the CM frame,
$\theta_{\ell \omega}^{\rm CM}$ ($\theta_{\ell K^{*0}}^{\rm CM}$), satisfy 
$\cos \theta_{\ell \omega}^{\rm CM} < 0.88$
($\cos \theta_{\ell K^{*0}}^{\rm CM} < 0.93$).
To remove two-photon background for the $e V^0$ modes,
we further require that the opening angle, $\alpha$, between the direction of
the total momentum of charged tracks and $\gamma$'s on the signal side
and that on the tag side satisfy the condition
$\cos \alpha > -0.999$ for the $e^- \phi$,
$\cos \alpha > -0.996$ for the $e^- \omega$, $e^- K^{*0} (\bar{K}^{*0})$ and
$\cos \alpha > -0.990$ for the $e^- \rho^0$ mode.

To identify signal $\tau$ decays,
we reconstruct the $\ell V^0$ invariant mass, $M_{\ell V^0}$, and
the energy difference in the CM frame, $\Delta E$, 
between the sum of energies on 
the signal side and the beam-energy, $E_{\rm beam}$.
Signal events should
concentrate around $M_{\ell V^0} = m_\tau$ and
$\Delta E = 0$, where $m_\tau$ is the nominal $\tau$ mass.
For the $\ell \omega$ modes, we used the beam-energy constrained mass, 
$M_{\rm bc}$, instead of the invariant mass $M_{\rm inv}$,
where $M_{\rm bc} = \sqrt{E_{\rm beam}^2 - (\vec{p}_\tau)^2}$,
in order to improve the mass resolution, which is smeared 
due to the $\gamma$ energy resolution.
In calculating the $\tau$ momentum $\vec{p}_\tau$,
we replace the magnitude of the $\pi^0$ momentum 
with the value obtained  
from the beam energy,
the energies of charged tracks on the signal side and
the $\pi^0$ direction measured by the calorimeter.
%the $\pi^0$ momentum the direction fixed.

The resolutions in $\Delta E$ and $M_{\ell V^0}$, 
evaluated using the signal MC, are summarized in Table~\ref{tbl:res}.
We define the signal region in the $\Delta E - M_{\ell V^0}$ plane
as a $\pm 3 \sigma$ ellipse.
In order to avoid biases in the event selection,
we blind the signal region until the analysis is finalized.

\begin{table}[htb]
\caption{Resolutions in $M_{\ell V^0}$ in MeV/$c^2$ and $\Delta E$ in MeV.
The superscripts low and high indicate the lower and higher sides of 
the peak, respectively.}
\label{tbl:res}
\begin{center}
\begin{tabular}{lcccc}
\hline
Mode & ~~~$\sigma_{M_{\ell V^0}}^{\rm high}$~~~ & ~~~$\sigma_{M_{\ell V^0}}^{\rm low}$~~~ & ~~~~$\sigma_{\Delta E}^{\rm high}$~~~~ & ~~~~$\sigma_{\Delta E}^{\rm low}$~~~~ \\
$\tau^-\to$ & & & & \\ \hline
$\mu^-\phi$   & $3.4\pm0.2$ & $3.4\pm0.2$ & $13.2\pm0.4$ & $14.0\pm0.5$ \\
$e^- \phi$   & $3.7\pm0.1$ & $3.6\pm0.1$ & $13.3\pm0.7$ & $15.4\pm0.7$ \\
$\mu^-\omega$ & $5.9\pm0.1$ & $6.2\pm0.1$ & $19.3\pm0.6$ & $30.3\pm0.8$ \\
$e^- \omega$ & $6.1\pm0.1$ & $6.5\pm0.1$ & $20.4\pm0.7$ & $32.5\pm1.3$ \\
$\mu^- K^{*0}$& $4.5\pm0.4$ & $4.5\pm0.4$ & $13.8\pm0.3$ & $14.4\pm0.4$ \\
$e^- K^{*0}$ & $4.3\pm0.1$ & $5.1\pm0.1$ & $12.9\pm0.3$ & $18.0\pm0.4$ \\
$\mu^- \bar{K}^{*0}$ & $4.7\pm0.1 $ & $4.4\pm0.1$ & $14.0\pm0.3$ & $15.0\pm0.3$ \\  
$e^-\bar{K}^{*0}$ & $4.6\pm0.1$ & $4.9\pm0.1$ & $12.6\pm0.6$ & $17.8\pm0.5$ \\
$\mu^-\rho^0$   & $5.6\pm0.1$ & $5.0\pm0.1$ & $13.9\pm0.4$ & $15.9\pm0.4$ \\
$e^- \rho^0$   & $4.7\pm0.1$ & $6.3\pm0.1$ & $14.5\pm0.4$ & $17.4\pm0.5$ \\
\hline
\end{tabular}
\end{center}
\end{table}

\section{Background Estimation}

\begin{figure}
\includegraphics[width=0.23\textwidth]{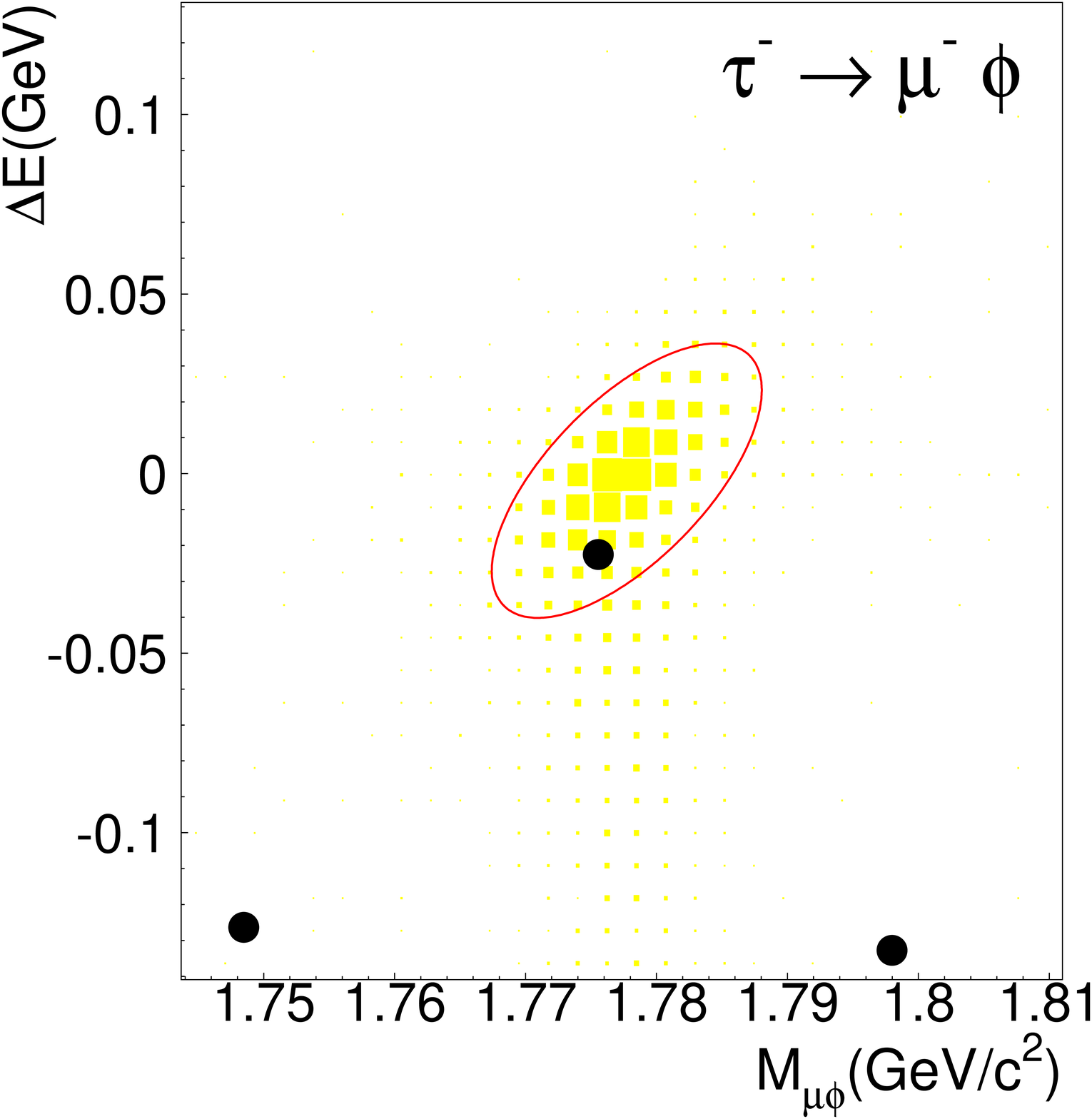}
\includegraphics[width=0.23\textwidth]{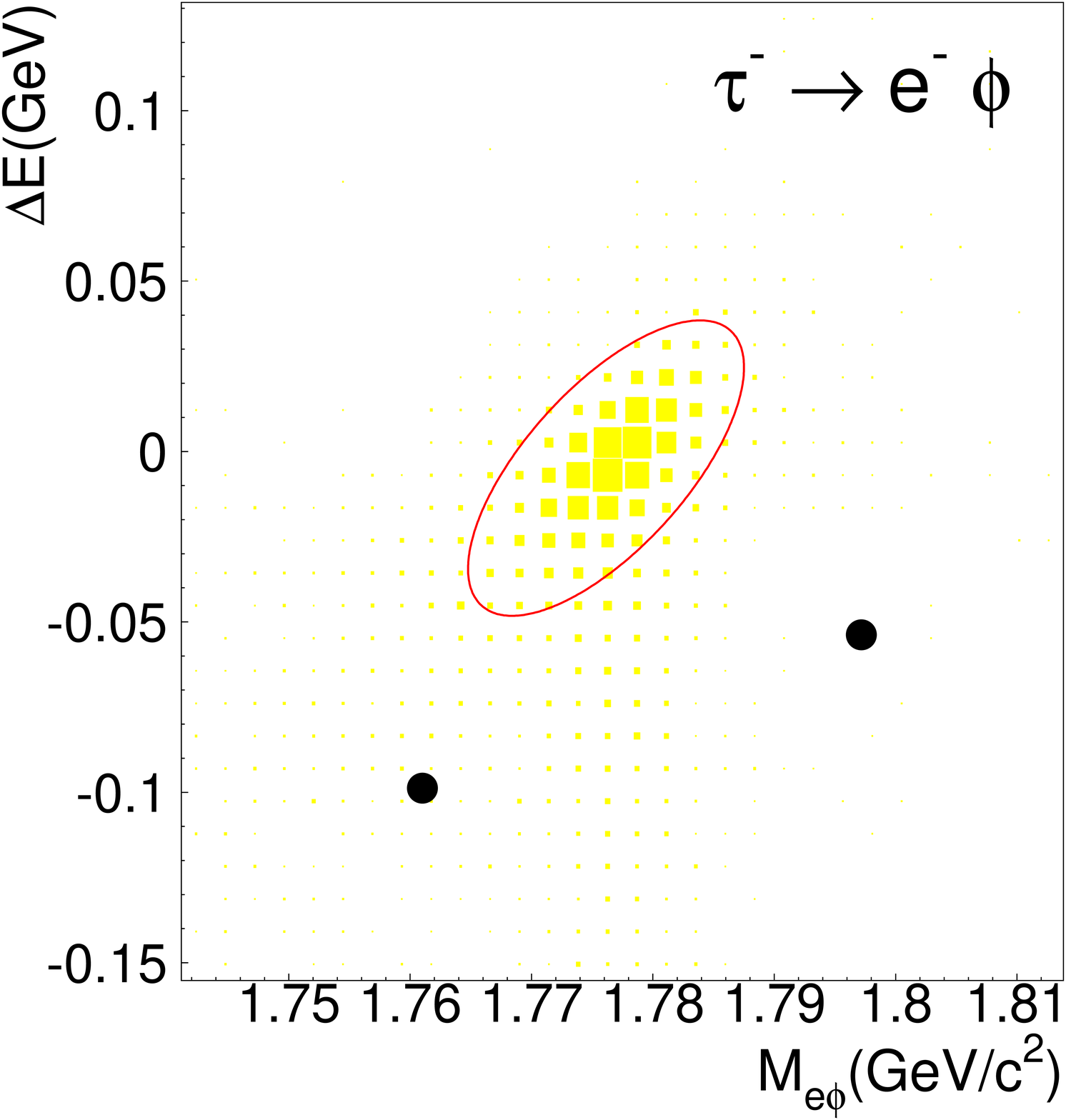}
\includegraphics[width=0.23\textwidth]{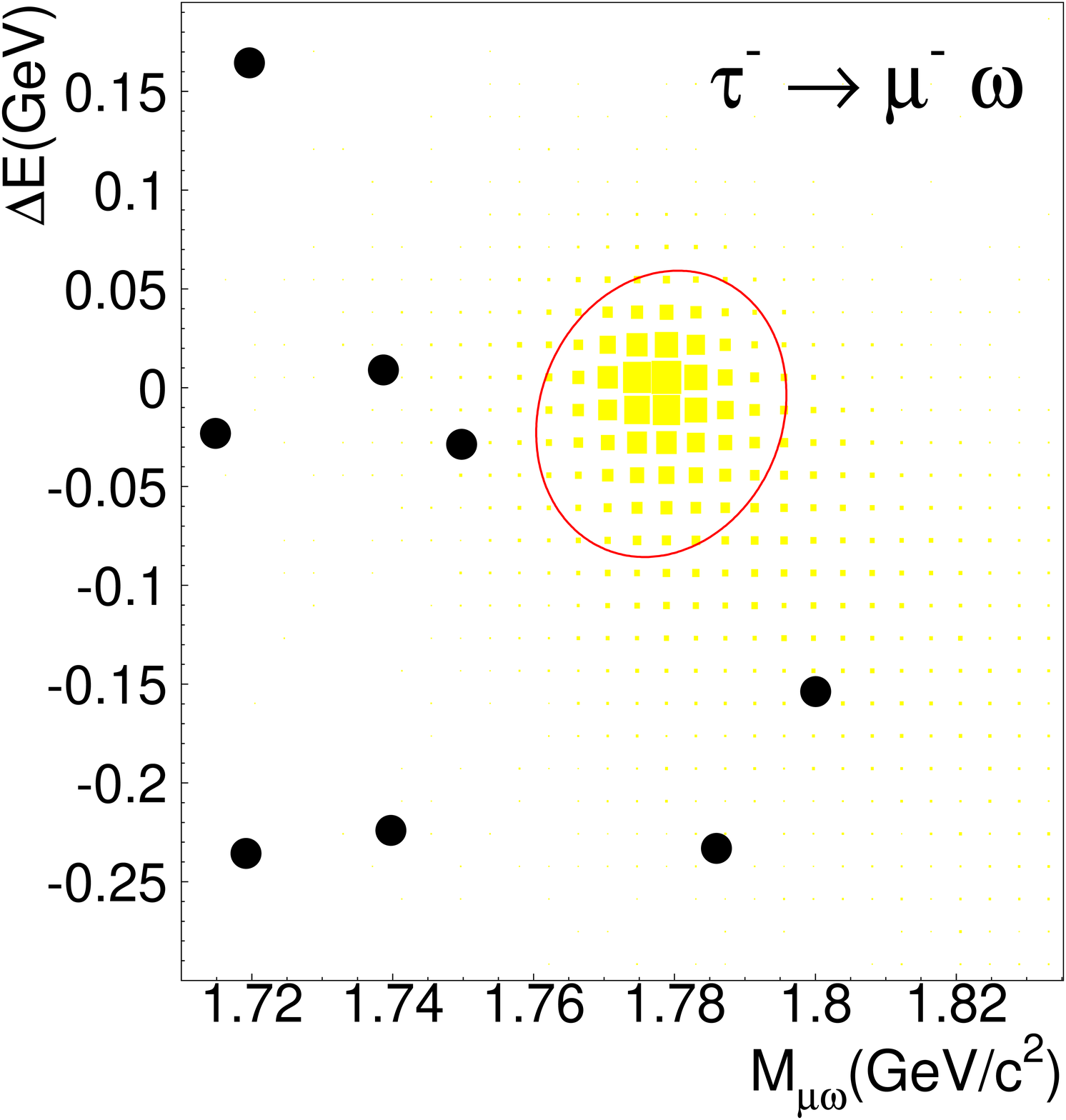}
\includegraphics[width=0.23\textwidth]{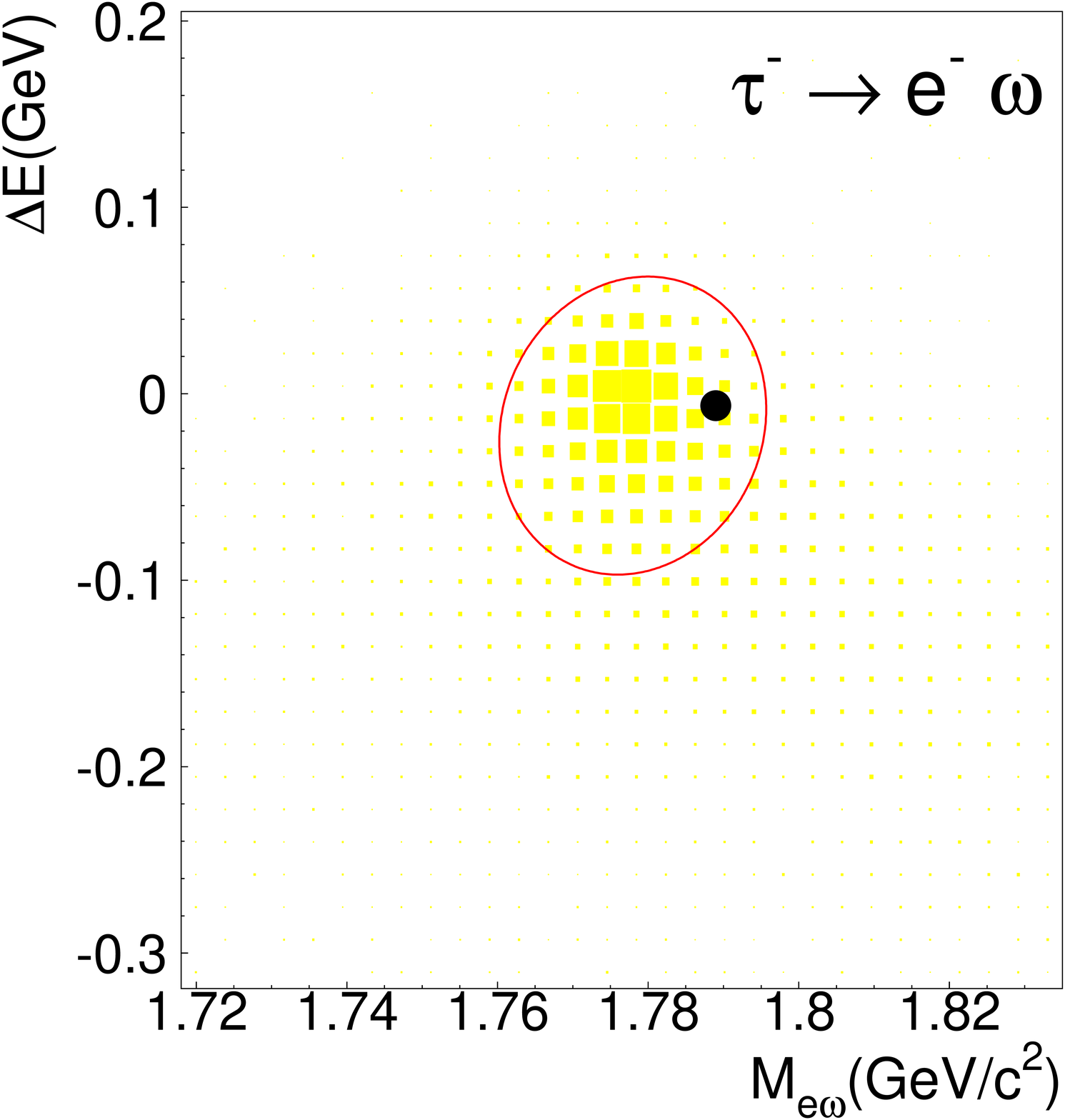}
\includegraphics[width=0.23\textwidth]{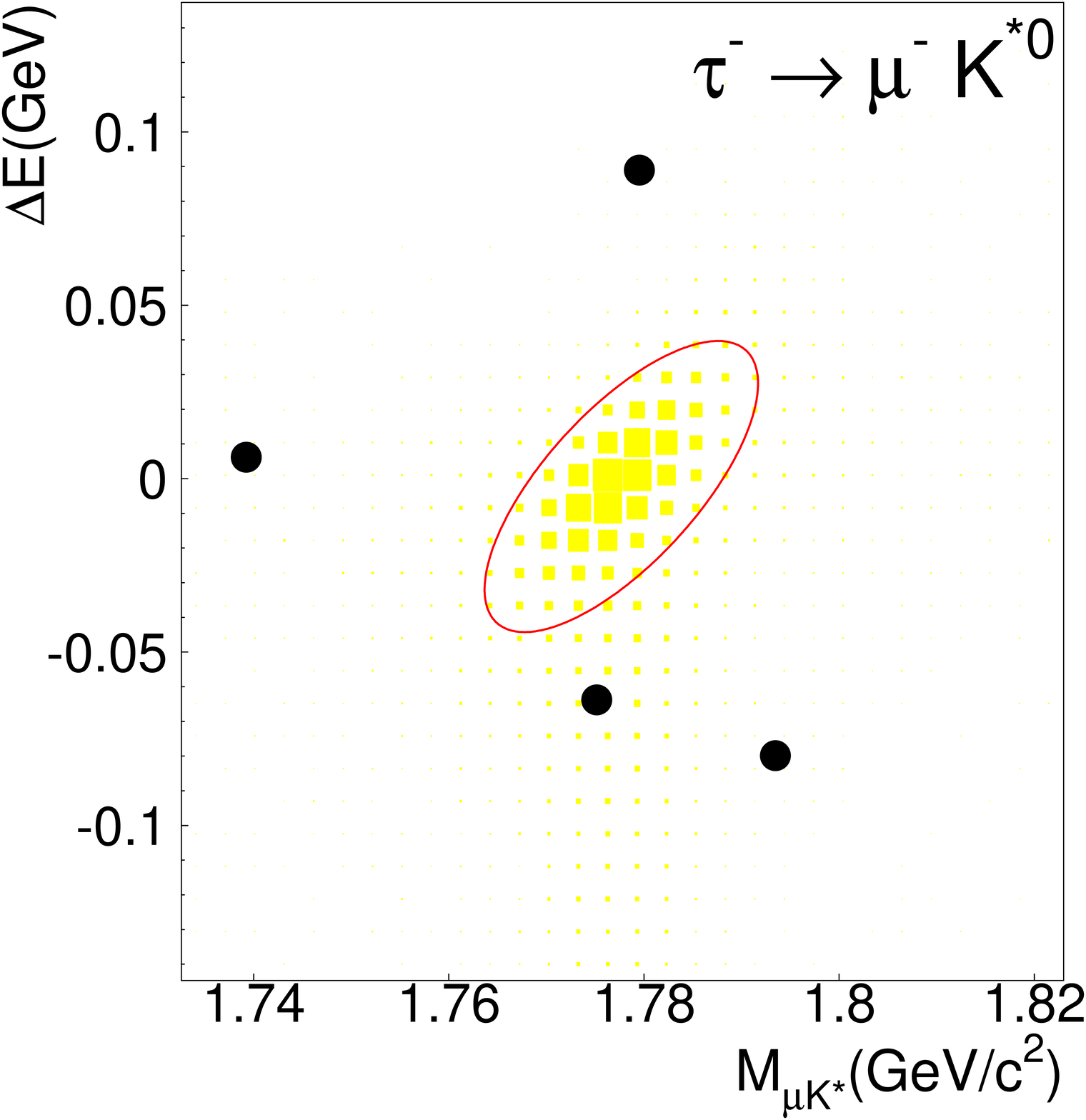}
\includegraphics[width=0.23\textwidth]{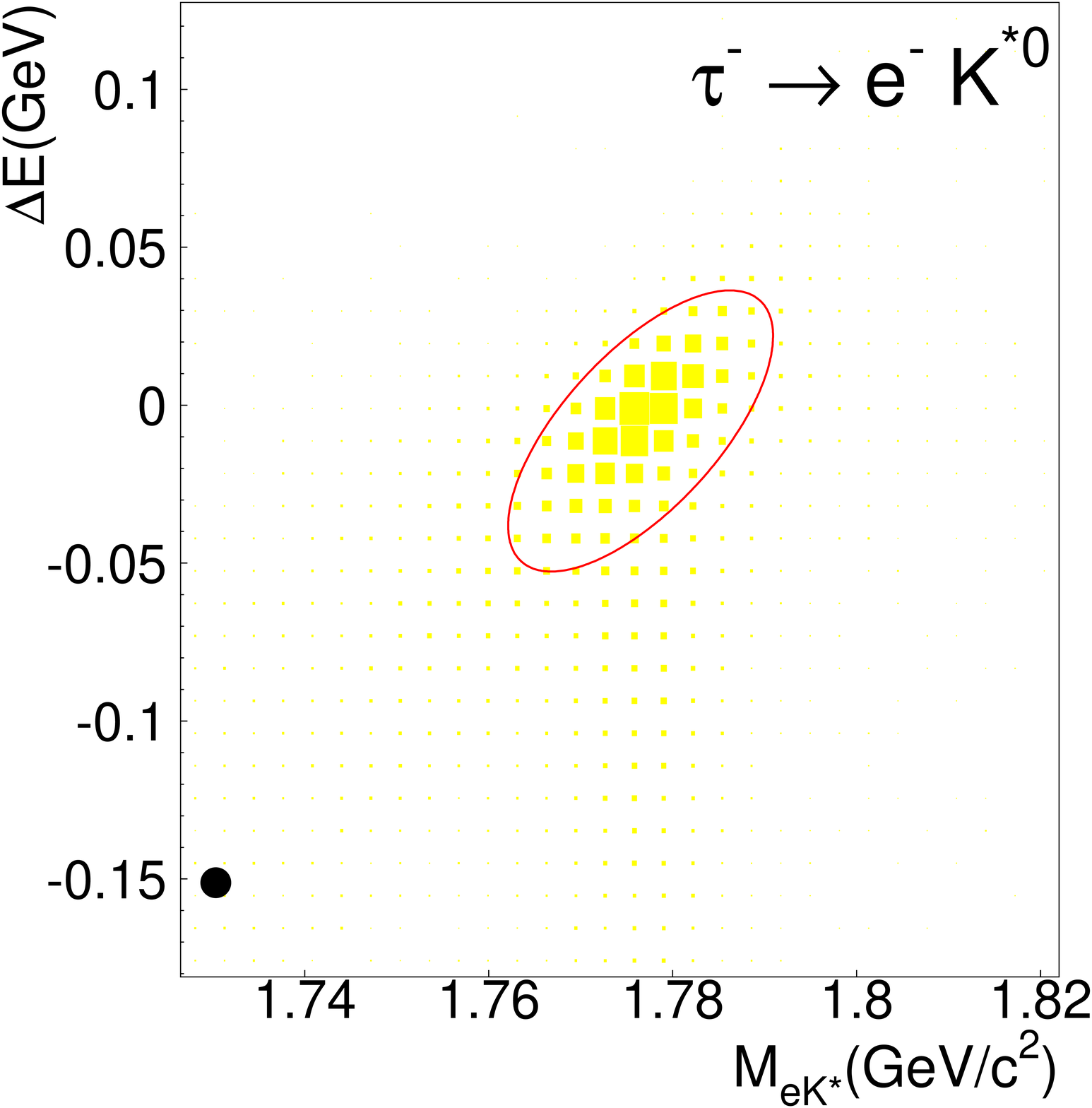}
\includegraphics[width=0.23\textwidth]{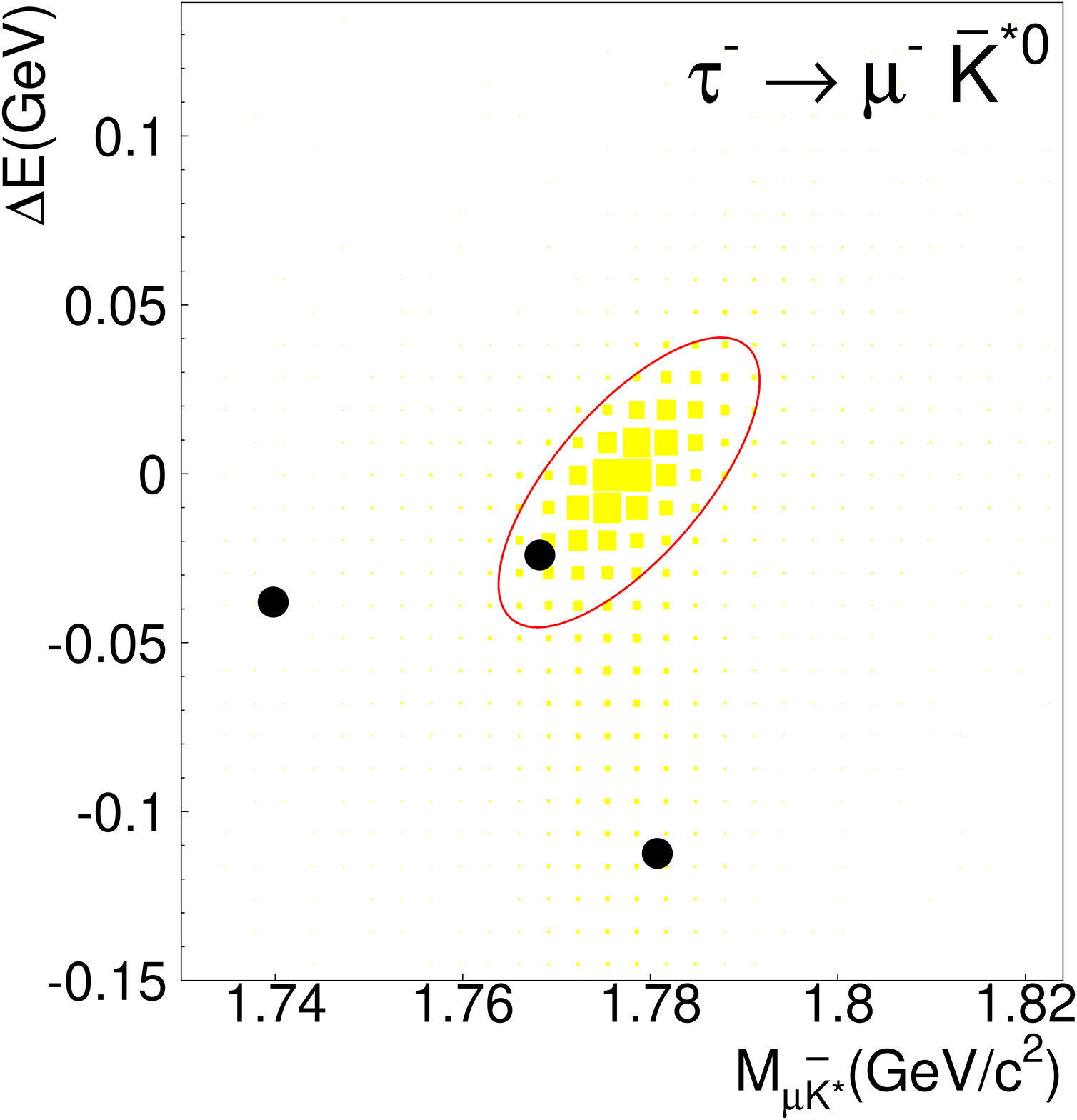}
\includegraphics[width=0.23\textwidth]{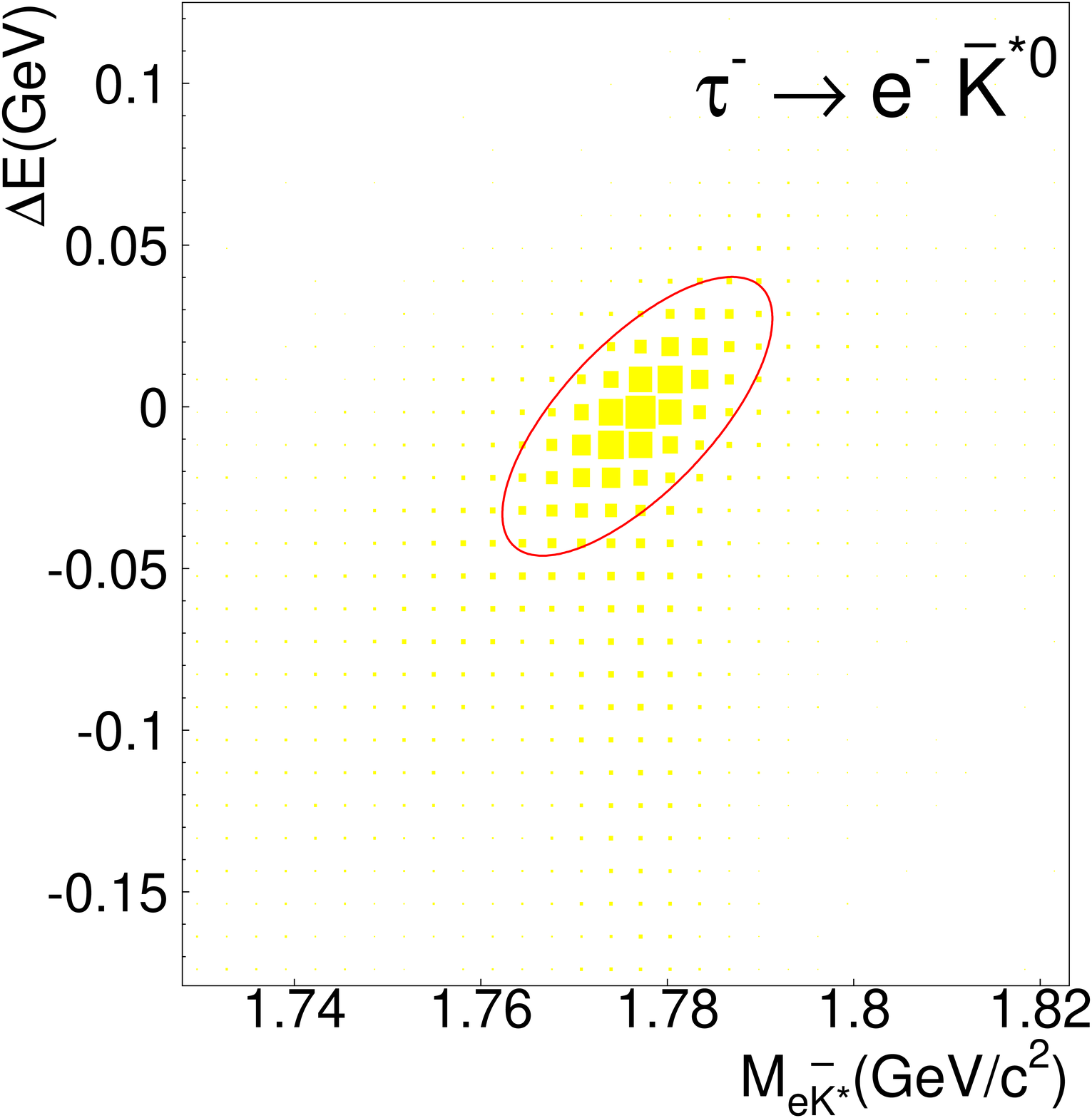}
\includegraphics[width=0.23\textwidth]{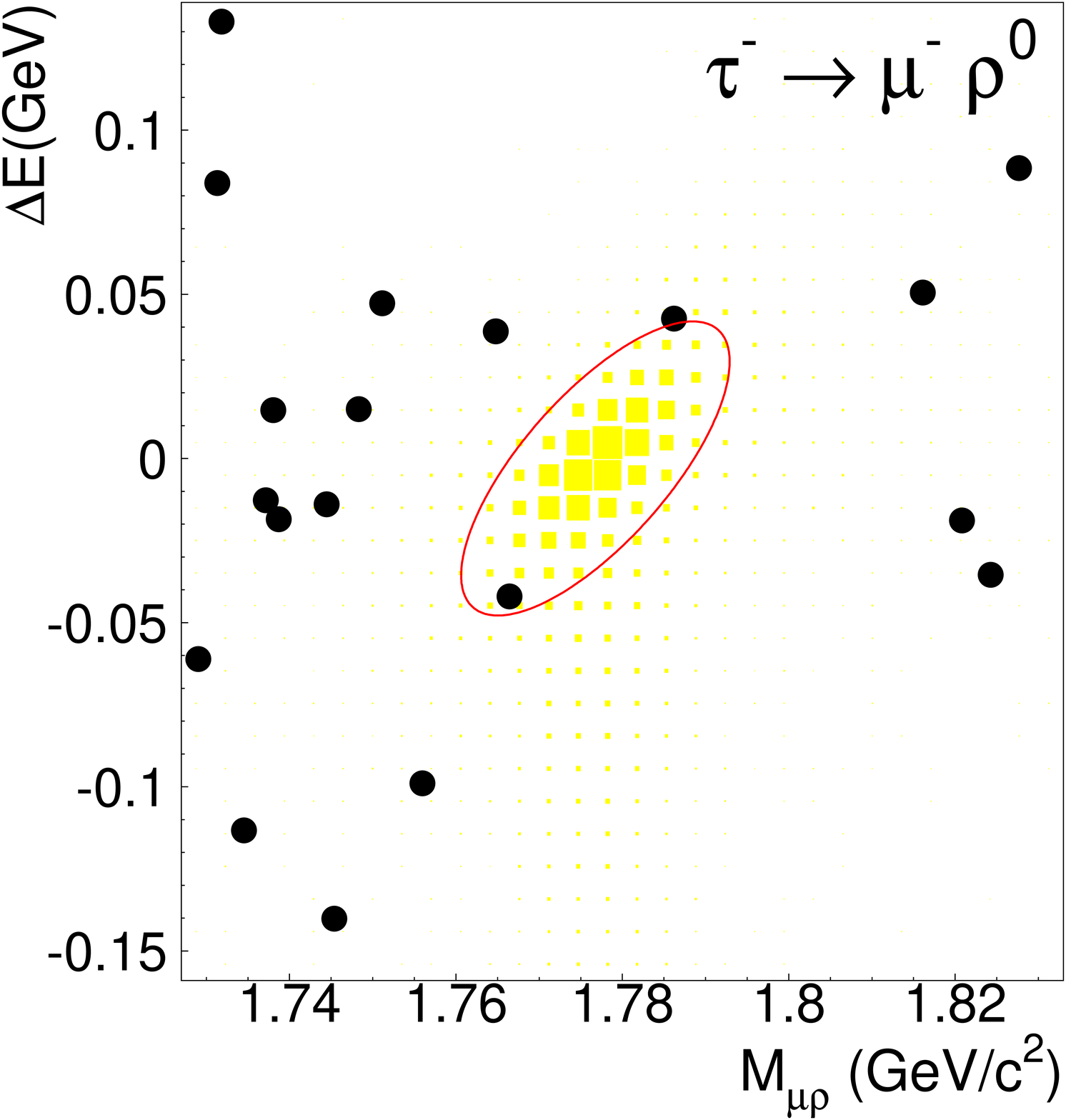}
\includegraphics[width=0.23\textwidth]{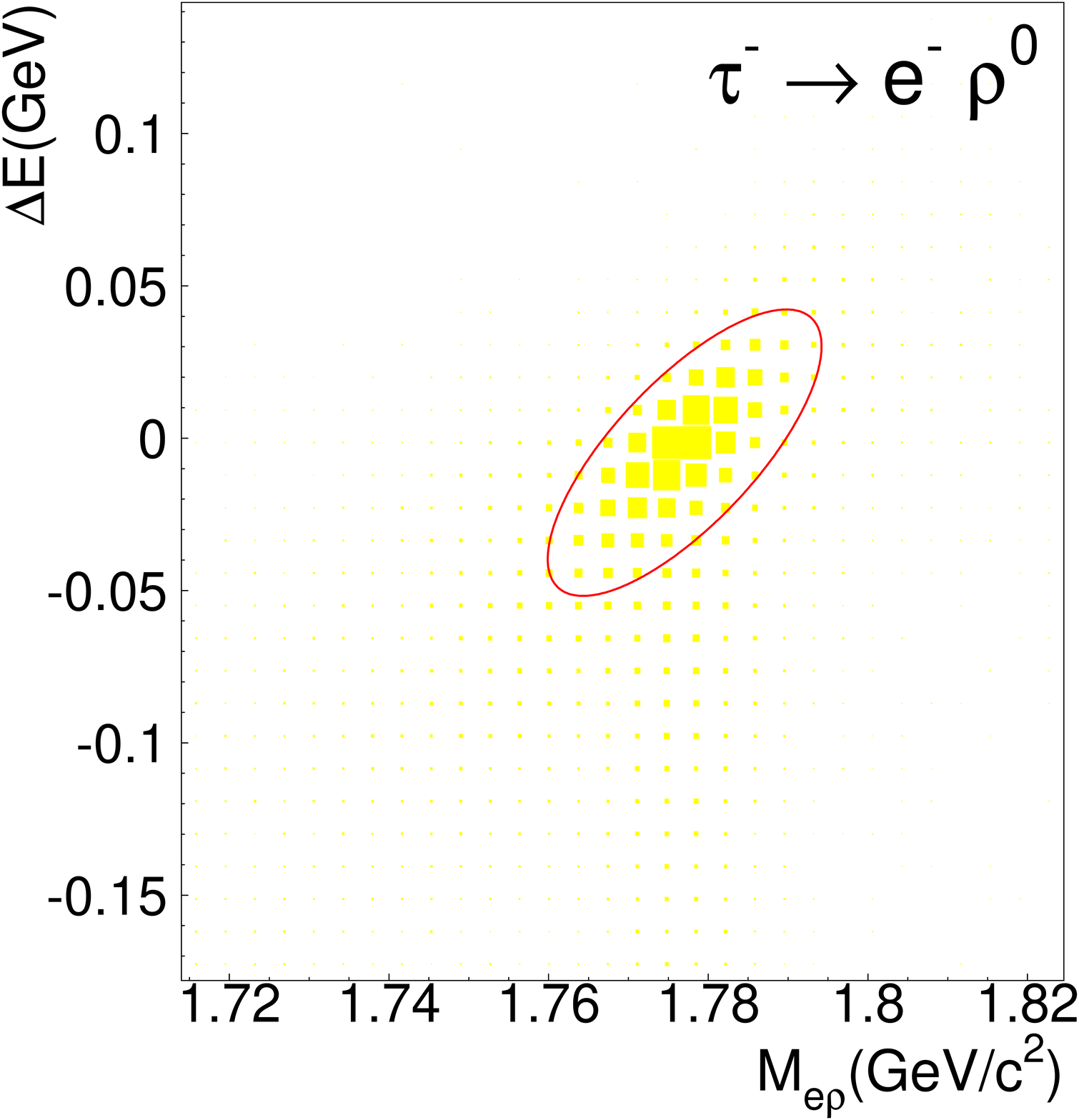}
\caption{Distributions of $\Delta E$ - $M_{\ell V^0}$ in the $\pm10\sigma$
box for 
(a) $\tau^- \to \mu^- \phi$, (b) $\tau^- \to e^- \phi$, 
(c) $\tau^- \to \mu^- \omega$, (d) $\tau^- \to e^- \omega$,
(e) $\tau^- \to \mu^- K^{*0}$, (f) $\tau^- \to e^- K^{*0}$,
(g) $\tau^- \to \mu^- \bar{K}^{*0}$, (h) $\tau^- \to e^- \bar{K}^{*0}$,
(i) $\tau^- \to \mu^- \rho^0$ and (j) $\tau^- \to e^- \rho^0$
after all selections. Dots are data and filled boxes show the signal MC.
%open circles show $\tau$-pair background MC and triangles are $q\bar{q}$ MC
The elliptical area is the $3\sigma$ signal region.}
\label{fig:result}
\end{figure}

\begin{table}[]
\caption{Number of events in the background region excluding the
signal region for data and MC.}
\label{tbl:bg1}
\begin{center}
\begin{tabular}{lcc}\hline
 Mode & \multicolumn{2}{c}{} \\
 $\tau^-\to$  & ~~~Data~~~  & MC \\
\hline 
 $\mu^-\phi$   & 2 & $1.72\pm1.23$ \\
 $e^-\phi$    & 2 & 0$\pm \genfrac{}{}{0pt}{}{1.19}{0}$               \\
 $\mu^-\omega$ & 7 & $10.46\pm1.91$  \\
 $e^-\omega$  & 0 & $1.07\pm0.62$ \\
 $\mu^- K^{*0}$ & 3 & $3.78\pm1.72$ \\
 $e^-  K^{*0}$ & 1 & $2.26\pm1.26$ \\
 $\mu^- \bar{K}^{*0}$ & 2 & $1.21\pm0.92$ \\
 $e^-  \bar{K}^{*0}$ & 0 & $0.31\pm0.31$ \\
 $\mu^-\rho^0$   & 12 & $4.82\pm1.25$ \\
 $e^-\rho^0$    & 0 &  $0.36\pm0.36$ \\
\hline
\end{tabular}
\end{center} 
\end{table}%

After all selections, a few events remain in the region
$-10\sigma^{\rm low}_{M_{\ell V^0}}<
M_{\ell V^0}<10\sigma^{\rm high}_{M_{\ell V^0}}$ and
$-10\sigma_{\Delta E}^{\rm low}<\Delta E<3\sigma_{\Delta E}^{\rm high}$,
which we define as a  background region. This region will 
be used to estimate the expected background
and is shown in Fig.~\ref{fig:result}.
Table \ref{tbl:bg1} lists the numbers of events in the
background region excluding the signal region.
The background is efficiently suppressed by the event selection.
The comparison between the data and MC shows reasonable agreement
for all modes, except for the $\tau^-\to\mu^-\rho^0$ mode.
For this mode one of the dominant background sources is
$q\bar{q}$, which at low multiplicity is poorly described by MC. 
For all other modes we estimate the number of
background events in the signal ellipse from the data in the background region
using the following method. 

Since for most of the modes there are single events 
only in the $\Delta{E}~-~M_{lV^0}$ plane both in the data and MC, 
we first study the background distribution in a sideband region 
larger than the background region 
($1.5$ GeV/$c^2<M_{\ell V^0}<1.95$ GeV/$c^2$ 
and $-0.5$ GeV$< \Delta E<0.5$ GeV) 
and find that the distribution of events in it is approximately flat 
in the background region.
Therefore we assume that this distribution is flat inside 
the background region and
estimate the expected number of background events in the signal
region from the number of data events
in the background region
times the ratio of the areas of the signal ellipse and background region.
If the number of data events
in the background region is zero,
we assign an upper limit of 2.44 events at the 90\% confidence level.
The expected number of background events obtained by this method is 
shown in the third column of Table~\ref{tbl:summary}. 

For the $\mu^- \omega$ mode, where the number of events is larger, 
we estimate the background 
contribution in the signal region using the shape of the background MC 
distribution normalized to the data yield in the sideband region of
the background region.

For the $\mu^- \rho^0$ mode, the background events in Fig.\ref{fig:result}
come mostly from $\tau^-\to \pi^-\pi^+\pi^-\nu$ decay 
and $q\bar{q}$ when one of the pions is misidentified as a muon. 
To estimate the background contribution we select a special event 
sample requiring $P_\mu<0.1$ for muon candidates instead of $P_\mu>0.95$.
The number of expected background events is then calculated
from the product of the number of events with $P_\mu<0.1$ in the signal region
and the muon fake rate.

\section{Results}

After unblinding the signal region single events only remain in some 
modes, see Fig.\ref{fig:result}.
The observed number of events in the signal region is consistent with 
the expected background. 
From the numbers of observed events in the signal region
and the numbers of expected background events, listed in 
the second and third columns of Table ~\ref{tbl:summary}, respectively,
we evaluate the upper limit on the number of signal events
at the 90\% CL, $s_{90}$, with systematic uncertainties
included in the Feldman-Cousins method~\cite{FeldmanCousins}
using the POLE code~\cite{HighlandCousins}.
In the cases when we give an upper limit of the expected background 
($\tau \to e^-\omega$, $e^-\bar{K}^*$ and $e^-\rho^0$ modes),
the number of background events is taken to be zero. 
This results in conservative upper limits.
The main systematic uncertainties on the detection efficiency come from 
track reconstruction (1.0\% per track), electron identification (2.2\%),
muon identification (2.0\%), 
kaon/pion separation (1.4\% for $\phi$ 
reconstruction, 1.1\% for $K^{*0}$ and 1.5\% for $\rho^0$),
$\pi^0$ reconstruction (4.0\%), statistics of the signal
MC (1.3\% for $\ell^- \phi$, 0.7\% for $\ell^- \omega$, 0.6\% for 
$\ell^- K^{*0}$
and $\ell^- \bar{K}^{*0}$, 0.5\% for $\mu^- \rho^0 $ and 0.6\% 
for $e^- \rho^0$)
and uncertainties in the branching fractions for $\phi \to K^+ K^-$
and $\omega \to \pi^+ \pi^- \pi^0$ (1.2\% and 0.8\%).
The uncertainty in the number of $\tau$-pair events mainly comes from the
luminosity measurement (1.6\%).

The upper limits on the branching fractions, ${\cal B}$, are calculated as
${\cal B} < \frac{s_{90}}{2 N_{\tau\tau} \epsilon}$,
where $N_{\tau\tau}=4.99 \times 10^8$,
is the total number of the $\tau$-pairs produced 
and $\epsilon$ is the signal efficiency including 
the branching fractions of 
$\phi \to K^+ K^-$, $\omega \to \pi^+ \pi^- \pi^0$,
$K^{*0} \to K^+ \pi^-$ and $\rho^0\to \pi^+\pi^-$~\cite{PDG}.
The resulting upper limits on the branching fractions are summarized in
Table~\ref{tbl:summary}.

\begin{table}[htb]
\caption{Summary of the number of observed events $N_{\rm obs}$,
the number of expected background events $N_{\rm exp}$,
detection efficiency $\epsilon$,
total systematic error $\Delta\epsilon/\epsilon$,
90\% CL upper limit of the number of signal events $s_{90}$
and 90\% CL upper limit of the branching fractions ${\cal B}$.}
\label{tbl:summary}
\begin{center}
\begin{tabular}{lcccccc}
\hline
Mode & $N_{\rm obs}$ & ~~$N_{\rm exp}$~~~ & ~~$\epsilon$~~ & ~~$\Delta\epsilon/\epsilon$~~ & ~~$s_{90}$~~ & ~~UL on ${\cal B}$~~ \\
$\tau^-\to$ & ~ & ~ & ~(\%) & (\%) & ~ & (90\% CL)\\
\hline
$\mu^-\phi$  & 1 & $0.17\pm0.12$ & ~3.14 & 5.2 & 4.17 & $1.3\times 10^{-7}$ \\
$e^- \phi$   & 0 & $0.18\pm0.12$ & ~3.10 & 5.3 & 2.27 & $7.3\times 10^{-8}$ \\
$\mu^-\omega$ & 0 & $0.19\pm0.20$ & ~2.51 & 6.3 & 2.22 & $8.9\times 10^{-8}$ \\
$e^- \omega$ & 1 & $<0.24$ &~2.46 & 6.3 & 4.34 & $1.8\times 10^{-7}$ \\
$\mu^- K^{*0}$& 0 & $0.26\pm0.15$ & ~3.71 & 4.8 & 2.20 & $5.9\times 10^{-8}$ \\
$e^- K^{*0}$ & 0 & $0.08\pm0.08$ & ~3.04 & 4.9 & 2.35 & $7.8\times 10^{-8}$ \\
$\mu^- \bar{K}^{*0}$ & 1 & $0.17\pm0.12$ & ~4.02 & 4.8 & 4.14 & $1.0\times 10^{-7}$ \\
$e^- \bar{K}^{*0}$  & 0 & $<0.17$ & ~3.21 & 4.9 & 2.45 & $7.7\times 10^{-8}$ \\
$\mu^-\rho^0$   & 1 & $1.04\pm0.28$ & ~4.89 & 4.9 & 3.34 & $6.8\times 10^{-8}$ \\
$e^- \rho^0$   & 0 & $<0.17$ & ~3.94 & 5.1 & 2.46 & $6.3\times 10^{-8}$ \\
\hline
\end{tabular}
\end{center}
\end{table}

\section{Summary}

We have searched for the LFV decays $\tau^- \to \ell^- \phi$, $\ell^- \omega$,
$\ell^- K^{*0}$, $\ell^- \bar{K}^{*0}$ and $\ell^- \rho^0$
using 543~fb$^{-1}$ of data obtained in the Belle experiment.
No evidence for a signal is observed, and the upper limits on the 
branching fractions
are set in the range $(5.9 - 18) \times 10^{-8}$ at the 90\% CL.
This analysis is the first search for the $\tau^- \to \ell^- \omega$ mode.
The results for the $\tau^- \to \ell^- \phi$,
$\ell^- K^{*0}$, $\ell^- \bar{K}^{*0}$ and $\ell^- \rho^0$ modes 
are $3 - 10$ times more
restrictive than our previous results obtained using 
158~fb$^{-1}$ of data.
The sensitivity improvement comes from a factor of 3.4-times larger statistics 
and an optimized analysis.
In particular, we have improved the conditions on 
$p_{\rm miss}$ and $m_{\rm miss}^2$ to reduce $\tau\tau$ and $q\bar{q}$ 
background as well as those on the opening angle $\alpha$ 
to reduce two-photon background.
As a result, better background suppression is achieved and the
efficiency is improved, e.g. it increases by a factor of
2.8 for the $\mu^-\phi$ and 2.5 for the $e^-\phi$ mode.
The new upper limits can be used to constrain the parameter space of 
various scenarios beyond the SM.

\section{Acknowledgements}

We thank the KEKB group for the excellent operation of the
accelerator, the KEK cryogenics group for the efficient
operation of the solenoid, and the KEK computer group and
the National Institute of Informatics for valuable computing
and Super-SINET network support. We acknowledge support from
the Ministry of Education, Culture, Sports, Science, and
Technology of Japan and the Japan Society for the Promotion
of Science; the Australian Research Council and the
Australian Department of Education, Science and Training;
the National Natural Science Foundation of China under
contract No.~10575109 and 10775142; the Department of
Science and Technology of India; 
the BK21 program of the Ministry of Education of Korea, 
the CHEP SRC program and Basic Research program 
(grant No.~R01-2005-000-10089-0) of the Korea Science and
Engineering Foundation, and the Pure Basic Research Group 
program of the Korea Research Foundation; 
the Polish State Committee for Scientific Research; 
%-> remove for now: under contract No.~2P03B 01324; 
the Ministry of Education and Science of the Russian
Federation and the Russian Federal Agency for Atomic Energy;
the Slovenian Research Agency;  the Swiss
National Science Foundation; the National Science Council
and the Ministry of Education of Taiwan; and the U.S.\
Department of Energy.


\begin{thebibliography}{99}
\bibitem{CC} Throughout this paper, 
the inclusion of the charge-conjugate decay modes is implied
unless otherwise stated.
\end{thebibliography}
\end{document}